# PhyNetLab: An IoT-Based Warehouse Testbed

Robert Falkenberg*, Mojtaba Masoudinejad†, Markus Buschhoff‡, Aswin Karthik Ramachandran Venkatapathy†,
Daniel Friesel‡, Michael ten Hompel†, Olaf Spinczyk‡ and Christian Wietfeld*
*Communication Networks Institute, †Chair of Materials Handling and Warehousing, ‡Embedded System Software Group
TU Dortmund University, Dortmund, Germany
Email:{robert.falkenberg, mojtaba.masoudinejad, markus.buschhoff, aswinkarthik.ramachandran,
daniel.friesel, michael.tenhompel, olaf.spinczyk, christian.wietfeld}@tu-dortmund.de

*Abstract*—Future warehouses will be made of modular embedded entities with communication ability and energy aware operation attached to the traditional materials handling and warehousing objects. This advancement is mainly to fulfill the flexibility and scalability needs of the emerging warehouses. However, it leads to a new layer of complexity during development and evaluation of such systems due to the multidisciplinarity in logistics, embedded systems, and wireless communications. Although each discipline provides theoretical approaches and simulations for these tasks, many issues are often discovered in a real deployment of the full system. In this paper we introduce *PhyNetLab* as a real scale warehouse testbed made of cyber physical objects (*PhyNodes*) developed for this type of application. The presented platform provides a possibility to check the industrial requirement of an IoT-based warehouse in addition to the typical wireless sensor networks tests. We describe the hardware and software components of the nodes in addition to the overall structure of the testbed. Finally, we will demonstrate the advantages of the testbed by evaluating the performance of the ETSI compliant radio channel access procedure for an IoT warehouse.

## I. INTRODUCTION

Flexibility of system, modularity and reliable throughput, in addition to the scalability play a major role in the quality of a materials handling and warehousing system [1]. Integration of embedded devices in the current systems is the first step into this direction that will enable classic warehouses evolving into decentralized modular systems with improved performance. There have been multiple research instances in the field of logistics which turned into implementation of distributed embedded systems for modern production, transportation and distribution strategies [2].

Although integration of single entities as standalone modules improve these structures, their full potential will be reached when these entities communicate with each other. Communication possibilities enables cooperative task fulfillment within a warehouse without the need for a central control unit storing all data and handling routing and order management.

Therefore, Cyber Physical Systems (CPS) are required instead of traditional operational units enhanced with embedded devices. These devices also known as *nodes* interconnect with each other to form a wireless sensor networks (WSN). Also, recent advancement in the field of low power wireless communication have made remote sensing applications possible and more efficient [3].

Smart connectivity to communicate with the available networks and using them for context-aware computation is an indispensable part of Internet of Things (IoT) [3]. Therefore, realization of these CPS objects in the field of materials handling is also a move in the overall direction of the IoT revolution [1].

Although the term IoT was first made in 1999 by Kevin Ashton in the context of supply chain management, the definition has been expanding into a wide range of applications during the past decade [3]. A diversity of fields brings multiple interpretations of this general concept. However, a simple explanation of this idea is "machine perception of the real world and interaction with it [4]." This perception can be of different real world entities; both physical or virtual things that interact with each other [5]. Atzori et al. [6] defines three paradigms for the realization of IoT as:

1) Internet oriented
2) Things oriented
3) Semantic oriented

The interdisciplinary nature of IoT requires this type of characterisation. However, this would be useful only in application domains with intersecting paradigms [3]. Warehousing application in a materials handling facility is exactly such a field which requires all three aspects of the IoT.

Wireless Sensor networks have gained a definite importance in the Industry 4.0 with the emerging concept of IoT [2]. Therefore, a complete understanding of WSN technologies is required before advancement into industry 4.0 applications.

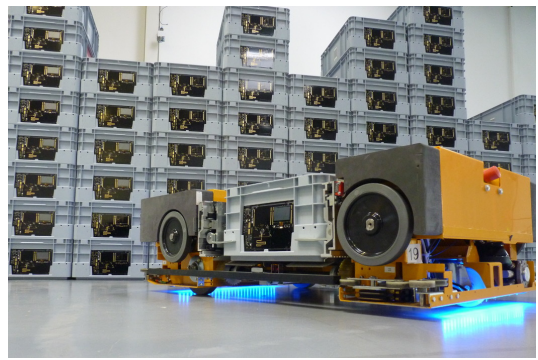

Fig. 1. Photograph of the PhyNetLab, a large logistics hall containing numerous smart containers, each attached with a communicating PhyNode. PhyNetLab serves as a large-scale testbed for the development of future, IoT-based warehouses.

Additionally, software development and development processes play an important role, as the code base for controlling a large swarm of devices needs to be maintainable, and implementing new features requires a well organized software milieu and a test environment to avoid disastrous mistakes.

Another challenge is imposed by radio communications. As bandwidth and range of radio connections is limited, a great mass of intelligent containers need to reduce and adapt their communication behavior accordingly. Furthermore, the channel access must be organized very efficiently to avoid collisions and the waste of scarce energy. As we will show later in this paper, commonly used approaches lead to a catastrophic increase in energy consumption in large-scale scenarios.

To tackle these challenges and enable the development, evaluation and validation of a real-life deployments, we present PhyNetLab (cf. Fig. 1), a large scale IoT testbed for future logistic systems, and perform an exemplary evaluation of an established channel access scheme.

## II. RELATED WORKS

WSN has been a major research topic during the last few years [7] which spans a diverse research area. From technical topics such as routing algorithms, energy harvesting to management, security and privacy.

According to this versatile range of work, wide variation of WSN testbeds are developed. In addition, there are federation of WSN combining multiple installations into a single wide spread platform. Some of the most famous platforms are:

*a) MoteLab:* A wireless network developed by Harvard and published in 2005 [8]. It was one of the first fully developed WSNs. MoteLab is an open source tool that is accessible on the Internet. Its web access facilitates remote programming and user scheduling.

*b) Future Internet of Things:* FIT/IoT-Lab [9] is a heterogeneous large-scale research facility with a federation of more than 6 locations and more than 2000 nodes altogether [10]. To interact with nodes it has command line interfaces through user virtual machines.

*c) Indriya:* is a low-cost 3D WSN testbed implemented at the National University of Singapore [11]. The 127 wireless nodes are connected with active USB cables. This USB infrastructure provides a back channel for remote programming and powering the sensor nodes.

*d) WISEBED:* is an IoT research facility with a heterogeneous implementation where each partner maintains its own testbed [12]. There is also an overlay network giving access to all testbeds as one, large IoT implementation for analysis. Anyhow, each testbed can also be handled separately [10].

In addition to these few examples, there are multiple WSNs with different sizes, number of nodes, communication and routing protocols. Some surveys, such as [4], [7], provide an overview of them. However, [7] categorize them according to their features into:

- general testbeds
- server based testbeds
- single PC based testbeds
- hybrid testbeds
- multiple site testbeds
- in-band management traffic testbeds
- specialized testbeds

PhyNetLab is considered to be a specialized test bed according to this categorisation because it is designed to be a wireless sensor network test bed built very close to a real world scenario where the major research challenges are to develop communication protocols and energy aware syntactics which are required for deep integration in an industrial system such as the materials handling facility.

## III. MOTIVATION

During the development of networks, validation of protocols is an essential step where PhyNetLab supports the modeling and testing of logical networks and protocols as well as sensor networks and their deployment. Most research teams use a theoretical analysis or simulation systems for this purpose. This is mainly because of the complexities occurring in embedded systems, electronics and energy measurements when empiric analyses are performed. In addition to hardware complexities, empiric modeling and testing is time consuming and costly [13], and it requires iterative reprogramming of dozens of nodes, locating them both physically and within the network topology [8].

However, both empiric and analytic models have issues to reproduce all influences of the actual sensor network environment. Although, both methods give an insight to the WSN, their lack of precision does not show the limitations of the network in a realistic scenario. It can be observed that, because of these limitations, more researches tend to rely on the real deployment of a network for evaluation and validation [13].

Also, there are some simulators such as Tossim or Cooja which are trying to bridge the simulation and experimentation gap. However, some real parameters of the system such as random parameters related to radio environment and system behavior such as failure are not simulated [13].

In addition to the above mentioned limitation of simulators, a platform's suitability can only be tested by including the physical constraints of the actual logistics entity under study [2]. This is mainly an issue while a platform's performance is dependent on the specific requirements of the material flow system. A material flow system has soft real-time aspects which are not directly dependent on the technical issues but the business process. The technical availability of a material flow system and its warehouse backbone has to maintain the desired throughput defined by the process [14].

The performance of embedded platforms is commonly rated by generic benchmarks. Although these benchmarks provide objective assessment objectives, diversity in the complex material handling applications makes benchmarking extremely hard if not impossible. Moreover, current planning approaches for the logistics application do not include wireless nodes integrated to the system [14]. Consequently, current tools are not adequate to solve existing problems with integration of WSN into the warehouse.

Furthermore, considering the industrial requirements of the IoT adds up another layer of complexity to a WSN. Efforts in this direction are summarized under the term *Industry 4.0* [2]. This term is a concept made of an embedded system attached to a physical entity, communicating with other CPS entities which can be other machines, networks or human beings.

With an industrial intent in mind for developing a testbed, the consideration of it fundamental elements is crucial. Hence, six principles of the testbed's architecture have to be considered which are [10]:

1) **Experimentation experience**: for energy harvesting, management and accounting to analyze energy neutrality of the system.
2) **Environment**: representing a real industrial environment and its dynamics.
3) **Availability**: of service for the performance analysis of the materials handling system.
4) **Scalability**: to modify the warehouse size according to dynamic business requirements.
5) **Heterogeneity**: for interoperation of multiple industrial devices and sites.
6) **Industrial impact**: to understand the effects of WSN integrated on a real warehousing process and its concern.

To analyze and evaluate IoT based warehouses and conquer the experimental limitations, PhyNetLab was developed as an IoT based warehouse testbed and is explained hereafter.

## IV. SYSTEM STRUCTURE AND HARDWARE COMPONENTS

Here, we describe the overview of the PhyNetLab, which is shown in Fig. 1. It is located in a large logistics hall providing space for a continuously extending amount of containers. Each container has an attached PhyNode at its front which enables communication capabilities with the infrastructure. The PhyNodes are associated to a base station, which is called Access Point (AP), as shown in Fig. 2. These APs are equally distributed in the PhyNetLab and serve as gateways to the internet. The intention of using multiple APs instead of a single gateway is reducing the necessary transmission power of the energy-constrained PhyNodes and increasing the network capacity by reusing the radio channels in each cell covered by distinct APs.

The test bed system architecture with network organization of data flow is listed below with a top down approach, where the top layer of the IoT platform has the servers hosted anywhere. This top layer is called the application layer. Followed by the application layer is the edge layer, where edge devices connect the servers in the application layer. The last layer is the operational layer or the IoT device layer. This is the layer where the end nodes are deployed and connected to various services in the application layer through the edge layer [10].

### A. Application Layer

In this test bed, the application layer is hosted and maintained within the project itself. There are different kinds of servers deployed in a micro-services architecture within the application layer. There is a database server for collecting and storing data and firmware management is performed using a version control management system such a GIT in a management server. An application programming interface is provided for clients within this network, that can be used for applications such as data analysis or coupling the experiment system with another system to emulate real-time scenarios within the research facility. In a real use case, this layer would also perform actions like resource restriction using access control levels, business logic and sending events to other systems using an event bus.

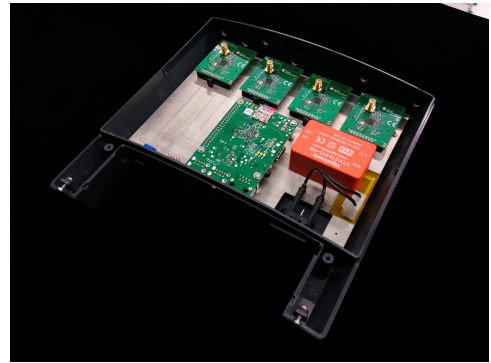

Fig. 2. Photograph of the AP which is serving as a gateway to the internet for the PhyNodes. Multiple APs are distributed in PhyNetLab to reduce the necessary transmission power of the PhyNodes, and allow a higher capacity by reusing the radio frequencies at different APs. I can be powered either by line voltage or by a battery to operate entirely wireless.

### B. Edge Layer

The edge layer is only a logical layer that provides a physical decoupling between the IoT device layer and the application layer. This layer consists of edge devices which become a proxy within the network to connect to the application servers. This also provides the flexibility of making local decisions in a physically colocated manner. The edge devices are called Access Points (APs) and based on embedded linux computers (for example: Raspberry Pi, Odroid) and also connect to low level devices using Serial Peripheral Interface (SPI) and or Universal asynchronous receiver/transmitter (UART) to receive data and administrate other IoT devices and networks. In addition, the APs are connected to the internet by an optimized Long Term Evolution (LTE) link [15] and can be powered by a battery for an entirely wireless operation. Hence, the deployment of APs does not require any static wiring in the hall.

The AP is connected to a CC1200 daughter board, which connects two CC1200 boards to the embedded computer using the SPI bus. In this case, the embedded computer runs the network stack for the Sub-1 GHz radio and only the physical layer and medium access in Sub-1 GHz area are performed using the CC1200 transceivers[1]. In addition, it contains two ZigBee capable CC2350 daughter boards, which operate in the 2.4 GHz frequency band [16]. They are also connected to the access-point computer but use a USB connection. On the motherboard of the AP an FTDI chip is used for every ZigBee transceiver board to convert a UART connection to USB. Only the necessary data is acquired from the ZigBee devices, since they run the network stack and send and receive data that are required for the application. The network layer messages can also be accessed when required from the APs. To make the AP scalable within a location, all the data in one AP is available on all other APs deployed in the hall using a distributed data storage system (for example: Memcached, Hazelcast or Redis). Mobility of the ZigBee devices can also be tracked and the data handover from one AP to another is made easier by deploying such an implementation. This layer also provides scalability of the whole test bed system by decoupling the physical location of the application layer, therefore any number of edge devices can be deployed after proper network capacity planning. They will be managed within a location locally using LTE or a local network together with a distributed data storage, or over the internet using the application layer management servers.

---
[1] http://www.ti.com/lit/ds/swrs123d/swrs123d.pdf

*C. IoT device layer - PhyNode*

There are two parts of the IoT device layer. One is the experiment platform and the other is a management platform. This management platform spans a ZigBee backbone network over the test bed which can be used for management of the end nodes. Therefore, the network is called Master Network and the board in the end nodes is called a Master Network board. The experimental platform is mechanically designed so that they can be replaced with another platform with a matching 8-pin connector. This provides flexibility on the platform. The swappable slave board (SSB) and the master network board (MNB) are described below with their individual hardware components.

*1) Master Network Board:* The main unit on the master network board (MNB) is the CC2530 integrated chip from Texas Instruments (TI). It provides health data, meta data, experiment firmware management for the experimental platform and other application end-points in ZigBee depending on the experiment scenario. It is a System on Chip (SoC) with a 2.4 GHz band transceiver that is IEEE 802.15.4 compliant with a Low-Power 8051 micro-controller architecture. The selection of such a RF transceiver is mainly due to its low power operation and a small spatial footprint in the printed circuit board (PCB). A Li-Polymer Battery with 2000 mAh capacity is used to power this section of the wireless node with battery protection using a BQ29700 chip that is integrated in the module. The MNB is connected using an 8-pin pole connector which powers the experimental platform when it needs to be operated using the battery. To prevent current leakage in the communication between the experimental board and MNB, isolation is required. It is provided using a simple voltage follower or unity gain amplifiers. An additional feature of the MNB is the high power infrared (IR) led. This IR led is used in the test bed implementation for validating various localisation algorithms and position tracing using a camera system which is not in the scope of this paper.

*2) Swappable Slave Board:* The heart of the swappable slave board is the MSP430FR5969, a ferroelectric random

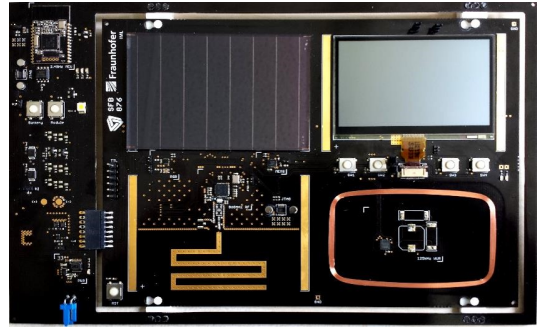

Fig. 3. Photography of the PhyNode, which is attached to every container in the PhyNetLab. It is equipped with a solar cell, display, radio interface, numerous sensors and a rechargeable battery.

access memory (FRAM) based MCU with a power consumption from $0.02\,\mu A$ in sleep mode up to $100\,\mu A$ in an active state. In comparison to a conventional flash RAM, the 64 kB FRAM in the MCU is very flexible concerning memory access and highly energy efficient. For RF communication the TI CC1200, a Sub-1 GHz transceiver with a 125 kHz wakeup receiver is implemented which communicates via an inductive coupling on the board and can react to a specified radio pattern. This pattern signal with works as a wake up source for the MCU that enables the MCU to switch between sleep mode and active mode. There are five onboard sensors with two MAXIM based ambient sensors for photovoltaic energy harvesting research. These sensors can measure RGB colour, infrared and ambient light. Two temperature sensors are used, one included in the MCU and one included in the MAXIM ambient sensor. Temperature is a critical measure since the photovoltaic cells also depend on an optimum operating range for energy harvesting. Furthermore a 3-axis accelerometer is also available.

The SSB can be programmed to work as an energy neutral device using energy harvesting. For this purpose and also for generic battery management, an ultra low power harvester IC with boost charger and autonomous power multiplexer is used. It is designed to allow switching to an alternative energy source if the solar charged energy storage is depleted. This measure is taken in order to make the hardware platform robust in terms of power and to easily understand and model its power requirements. For power distribution a low quiescent current, multi-mode, highly flexible power management integrated circuit (PMIC) is chosen. Finally, interaction with humans is possible using four buttons and a WQVGA (400x240px) memory LCD. The power source for the LCD can be cut off depending on the use case of the device, so that it does not impact the overall power consumption when it is disabled this way.

## V. SOFTWARE COMPONENTS

*A. Kratos - IoT End Device Operating System*

To enable a rapid development of IoT applications in ultra low-power systems, we developed Kratos[2]. Kratos supports ap-plication developers with a comprehensive set of C++ library functions that can be "tailored" to suit the needs of a distinct application and thus can save resources by only deploying the required system components.

---
[2]Kratos, the god of power, and a recursive acronym for "Kratos, a Resource Aware, Tailored Operating System".

To support the modularization that is necessary to keep an operating system highly configurable, we used AspectC++[3], a set of language extensions to facilitate aspect-oriented programming concepts into C and C++ [17], [18]. Several studies have shown the benefits of using AspectC++ for modularization and configurability [19]–[21].

Kratos delivers interfaces to different system architectures and features a set of useful system properties like interrupt synchronization, preemptive thread scheduling, peripheral driver interfaces and power management mechanisms.

Kratos was mainly developed with PhyNetLab in mind, and hence fully supports the PhyNode platform. Furthermore, PhyNetLab and Kratos allow mutual research on hardware and software design for large-scale and energy-aware IoT deployments. By that, PhyNetLab gives us insights on how to design energy and network-aware software components for an operating system under heavy resource constraints. In the other direction, software requirements for enabling maintainability and the deployment of applications for PhyNetLab led to choices and design decisions concerning the PhyNetLab components. As an example, the selection of optimal radio transceivers was narrowed down by their interface abilities, so that an energy aware driver implementations is supported. E.g., transceivers that required special, software driven handshake mechanisms were sorted out, as these transceivers require the CPU to be active permanently during transfers. Better fitting transceivers could be handled by pure hardware mechanisms (like the MSP430's direct memory address feature for bus transmissions), thus heavily reducing CPU load.

The main focus of Kratos is on power management. Here, we developed methods to incorporate detailed energy models of all components of a system into the system drivers. These energy models can be derived automatically by using power measurement techniques in a loop. This enables us track the power consumption of all nodes over time and drive application decisions on this knowledge. We elaborate this feature in greater detail in Section VII, where we determine the energy models for an experimental setup shown in Section VIII.

We postpone the presentation of the comprehensive details about the novel design principles of Kratos to a future work. Additionally, the Kratos source-code will be freely available as open source as soon as we consider it mature enough for the use by external users.

### B. Edge Device software

The edge devices serve as central gateways for PhyNode management. Every single AP keeps track of the available PhyNodes via device discovery and dynamic addressing. Additionally, it can manage the goods stored in the warehouse by communicating with the smart containers. Both queries, either to specific containers or via broadcast polls, and updates for the item type and count in a specific container are supported.

[3]http://www.aspectc.org/

TABLE I
PROPERTIES OF THE RADIO INTERFACE

| Physical Properties | |
|---|---|
| Device | CC1200 |
| Frequency | 866 MHz (SRD Band) |
| Maximum Transmission Power | 25 mW (14 dBm) |
| Receive Filter Bandwidth | 104 kHz |
| Modulation | 2-GFSK |
| Deviation | 20 kHz |
| Bit Rate | 38.4 kbit/s |
| Duplex Mode | Time Division Duplex |
| **Protocol Properties** | |
| Channel Access | LBT with random backoff |
| Packet Size (Payload) | Dynamic 0 B to 126 B |
| Address Width | 8 bit |

To allow reliable communication, a combination of device address management and protocol sequence numbers is used. Each request contains a sequence number, which must be echoed back by the addressed clients. When addressing a specific client, this sequence number can be used to determine whether a reply was received and, if that is not the case, retransmit the original request.

However, this method is not used for broadcast requests such as polling for goods in the warehouse containers. In this case, only the containers which contain the requested good send a reply. As the edge devices have no complete view of the warehouse contents, they cannot calculate the number of expected replies and hence cannot detect lost packets. Sending individual requests to each PhyNode would solve this issue, but at the cost of a poll time which grows linearly with the number of warehouse containers. At rates of thousands of containers in a single warehouse, this is not feasible.

## VI. COMMUNICATION

In order to add wireless communication capabilities, the PhyNode is equipped with a Texas Instruments CC1200 low-power, high-performance transceiver. The device is highly configurable in terms of e.g. modulation, power consumption, data rate, channel access and interrupt signals. Therefore it allows an exploration of a large design space to find the optimal configuration for the given scenario.

In this paper, an exemplary radio driver implementation configures the interface as shown in Tab. I. It operates in the Short Range Devices (SRD) band at 866 MHz with a bit rate of 38.4 kbit/s.

The expected high number of devices in the warehouse requires a collision-avoiding channel access mechanism. Otherwise, in case of an uncontrolled random access, the throughput over the air interface would achieve at maximum 18 % because of collisions [22], [23]. Since collisions require a retransmission of the data, which might collide again as well, this would waste a large amount of scarce energy. In addition, multiple replies to a single broadcast (or multicast) message

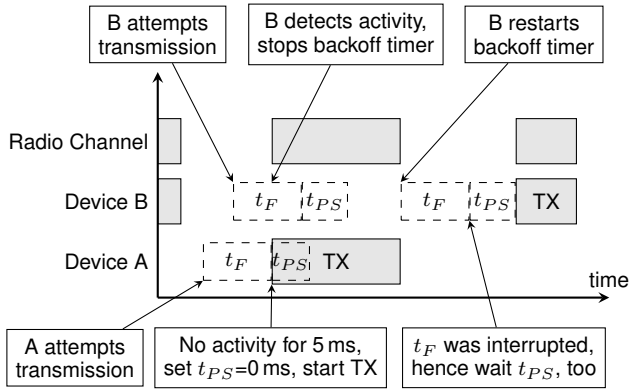

Fig. 4. Example for the applied LBT algorithm in the testbed. The dashed blocks represent back-off times while solid blocks represent actual radio transmissions over the air.

might always produce a collision, in case of an equal message-processing time.

Therefore, we implemented a Listen Before Talk (LBT) channel assessment in our setup, which conforms to the ETSI SRD standard [24] and complies with the regulatory constraints of that band. Instead of just accessing the channel for the transmission of a new data packet, the transceiver first listens on the channel for a total back-off time $t_L = t_F + t_{PS}$, where $t_F$ is fixed to $5\,\mathrm{ms}$ and $t_{PS}$ is a randomly chosen between $0\,\mathrm{ms}$ and $5\,\mathrm{ms}$. An example of the described channel access procedure is shown in Fig. 4. If DEVICE A with a pending transmission detects no radio activity throughout $t_F$, it shall subsequently perform its transmission by resetting its $t_{PS}$ to zero. However, if the transceiver detects any radio activity during $t_F$ (DEVICE B), the back-off timer is halted and will be restarted, when the channel is free again. In this case $t_{PS}$ will not be reset to zero. Instead, the device must back-off for the full total period $t_L$.

The response of broadcast messages requires a special handling: After receiving a broadcast message, multiple devices may attempt to reply simultaneously. Although those devices back-off for the period $t_F$, their messages will collide if no other activity happens on the channel during this period. Therefore, an additional random back-off of $0\,\mathrm{ms}$ to $5\,\mathrm{ms}$ is added in advance to each broadcast reply.

Fig. 5 shows a spectrogram of the radio channel captured by a real-time spectrum analyzer in the operating band. The dark blue background represents a free spectrum and background noise, while lighter colors show actual radio transmissions. In the figure three devices are attempting a transmission while the channel is occupied by a narrow jamming signal. The jamming signal is not part of the actual protocol, but is used here to trigger a clear channel assessment of multiple devices at the same time. As soon as the signal disappears, all devices simultaneously start the clear channel assessment algorithm. After $6\,\mathrm{ms}$, which is $t_F = 5\,\mathrm{ms}$ and $t_{PS} = 1\,\mathrm{ms}$, the first devices accesses the channel an performs its transmission. The remaining two devices, detect this transmission and restart their channel assessment as soon as the channel is free again.

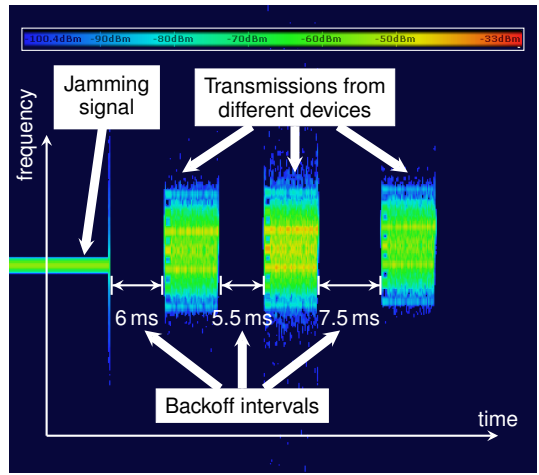

Fig. 5. Spectrogram of the radio channel captured by a real-time spectrum analyzer. Three devices transmit a data packet with respect to the LBT algorithm and the randomly chosen back-off time.

Subsequently, the next device transmits its packet after a back-off period of $t_F = 5\,\mathrm{ms}$ and $t_{PS} = 0.5\,\mathrm{ms}$, and finally the last device after $t_F = 5\,\mathrm{ms}$ and $t_{PS} = 2.5\,\mathrm{ms}$.

In order to reduce the power consumption of the transceiver during idle times, where the devices waits for an incoming transmission, we implemented a low power listening mode into the device driver. Instead of continuously listening on the radio channel, the transceiver is put into a low power sleep mode and wakes up in intervals of $4.7\,\mathrm{ms}$ for $0.2\,\mathrm{ms}$ and sniffs for a packet on the channel. This reduces the idle power draw from $23\,\mathrm{mA}$ in continuous RX to $1.5\,\mathrm{mA}$ in low power listening mode. However, this approach requires the transmission of extended packet preambles in a length of the sleep interval to prevent the miss of packets. Since transmissions of packets are generally less frequent than wake up events, this approach is energy-efficient in any case. The low power listen mechanism is disabled during a pending transmission in order to keep track of the exact moment, where an active transmission on the channel ends. This ensures the conformance to the previously described ETSI clear channel assessment.

## VII. ENERGY ACCOUNTING

To facilitate the accounting of consumed energy in each node, we use a deterministic finite automata (DFA) model that can be created for each hardware component of a system as described in [25]. The Kratos operating system was built with such models in mind and supports incorporating DFA models into the driver layer for peripheral components. To do so, it uses AspectC++ features (see Sec. V) to weave energy accounting program code into the respective driver functions that correspond to state transitions in the DFA model. We use this feature for the CC1200 radio transceiver as the main power consumer in the system. Thus, on each utilization, the driver keeps track of the transceiver's energy consumption and the total energy usage can be queried at the end of an experiment.

The energy model consists of the states RX, TX and IDLE and has one transition for each driver function which communicates with the radio, e.g. `send(...)`, `listen()` or the packet-received interrupt. Each state is annotated with its average power consumption $P_{state}$ and each transition with the transition energy $E_{tran}$. Combined with the time since the last function call $t_{state}$, this allows energy accounting by adding $P_{state} \cdot t_{state} + E_{tran}$ to the energy counter after each transition. The corresponding DFA is shown in Fig. 6.

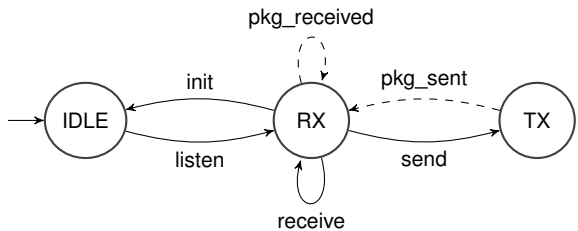

Fig. 6. DFA model for the CC1200 radio module in conjunction with the exemplary driver. Dashed transitions are initiated by the radio and signaled via interrupts.

To determine the cost of each state and transition, we use a toolset for automated energy measurements which is delivered by Kratos. Here, each state and transition of the radio module is traversed in a benchmark application, and driver calls are logged via UART and digital outputs. The application is automatically generated based on the DFA, while the log output allows automatic extraction of state power and transition energy data, which is in turn used for energy accounting by weaving it into the driver code.

The accounting code consist of variables for the current radio state, the time of the last state change and the total energy used which are updated after each transition. At an expected average power draw of up to 100 mW, our approach can log up to five years of high-precision energy data at the cost of 24 µs accounting overhead per modeled function call.

## VIII. APPLICATION EXAMPLE

In order to bring out the benefits of the proposed PhyNet-Lab, we present an evaluation of a typical application in future warehouses. Within such warehouses, incoming orders might autonomously send broadcast messages into the network to poll for particular goods. These messages will be replied by only those PhyNodes, which contain the requested product, and inform the inquirer about the contained quantity. He then selects a subset of PhyNodes, which fulfill the requested quantity, and requests to the network a delivery of those goods.

In this use case arises questions about radio channel congestion, reliability and energy consumption of the PhyNodes when replying to those polls simultaneously. To answer these questions, we performed real field measurements in the PhyNetLab (cf. Fig. 7) and set up a network of 38 PhyNodes attached to an AP. The nodes were distributed in the PhyNetLab in such a manner, that each node can perceive the activity of any other PhyNode in the network. Therefore, the free channel assessment algorithm is not negatively influenced by hidden stations. Each measurement run consists of assigning the same product to a subset of the attached PhyNodes and ten polls for that product by the AP. As soon as a PhyNode receives a matching poll message for the assigned product, it replies the call with its address and the contained amount with respect to the clear channel assessment procedure described in Sec. VI. The remaining PhyNodes perform no transmissions during the run. All participants in the network count the number of successfully received and sent messages on their side and transmit those statistics to the AP after each run. The statistics also include the accounted energy of the radio transceiver, as explained in Sec. VII. For comparability, each run comprises an interval of 11.75 s, hence does not depend on the number of exchanged messages. The interval in encapsulated by broadcast messages for starting and stopping a run. These messages are required, because statistics and setup messages of the PhyNodes are exchanged over the same link as the performance runs but must be excluded from statistics in this series of measurements. Therefore, each PhyNode resets its statistics and accounted energy when receiving a start message and freezes the values after receiving a stop message.

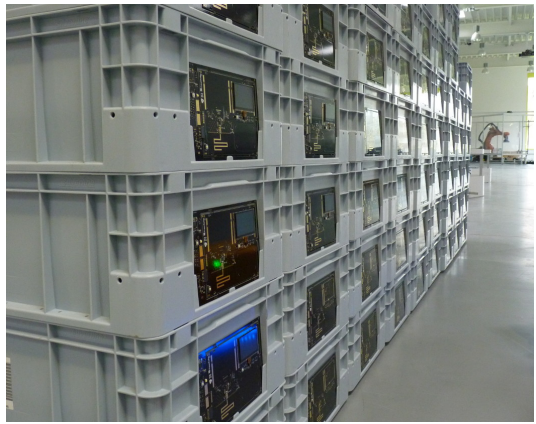

Fig. 7. Measurement setup in the PhyNetLab with 38 active PhyNodes for the evaluation of energy consumption and transmission reliability in congested scenarios.

## IX. EVALUATION AND RESULTS

In this section we evaluate the measurements from the application example in Sec. VIII, which challenged the ETSI collision avoidance algorithm of the radio interface during product polls to the warehouse. These product polls, which are transmitted as broadcasts by the AP into the network, trigger instantly numerous PhyNodes to transmit a reply message. Fig. 8 shows the achieved packet throughput $T$ during the measurements. It is defined as

$$T = \frac{N_{RX}}{\sum_{A} N_{TX}}, \qquad (1)$$

where $N_{RX}$ is the number of successfully received messages by the AP and $N_{TX}$ is the number of transmitted packets by each PhyNode from the active set $A$.

The algorithm performs very efficient for low number of concurring devices and enables a throughput above 80 % in case of 8 or less devices. In the range of 1 to 17 devices the throughput behaves nearly linearly and undercuts 50 % in case of 17 devices. Higher numbers of simultaneous attempts have only a negligible effect on the throughput, which stays nearly constant at a level of 50 %.

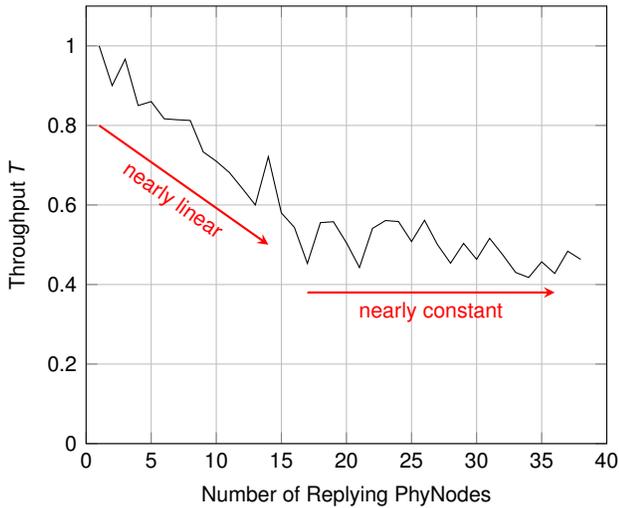 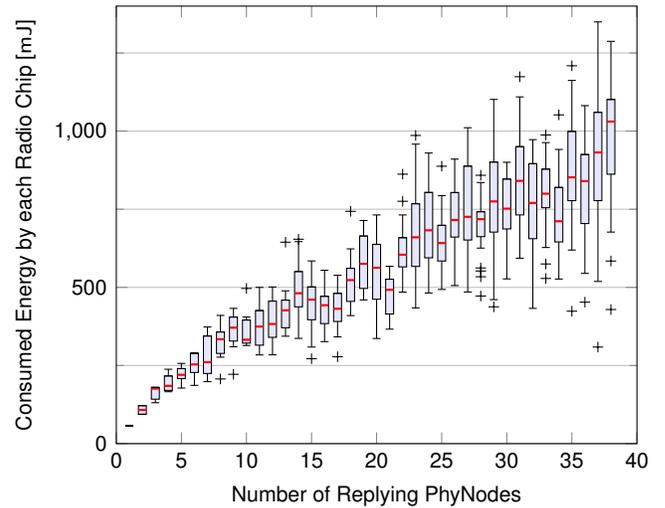

Fig. 8. Performance of the ETSI SRD collision avoidance algorithm in case of massive simultaneous replies to a broadcast message (poll for a particular product in the warehouse). With an increasing number of concurring transmission attempts, the packet throughput decreases due to more collisions.

Fig. 9. Energy consumption of the radio interfaces of active PhyNodes in the presented scenario. A single sample represents the total energy consumption of a PhyNode for receiving 11 packets and transmitting 10 replies in a constant interval of 11.75 s.

While this degree of throughput is typically not acceptable in a common communication system, it is still reasonable in the addressed logistics scenario. Considering, i.e., an incoming order for a particular product stored in the warehouse, where numerous containers comprise many entities of the demanded product, it is not necessarily required to receive all replies. Instead, it might be sufficient that *enough* containers reply to satisfy the demanded amount. Otherwise the poll could be repeated and uncover further containers, which suffered of a collision in the first place.

More critical is the impact of congestions on the energy consumption of the PhyNodes, which is shown in Fig. 9. In case of a single replying device, which reflects the minimum required energy for a test run, the radio interface consumes in average 57 mJ. By adding one additional device, the average energy consumption of each device more than doubles to 123 mJ. This is caused by a significantly higher amount of time spent in active receive mode: A device needs to wait in receive mode until a preceding transmission finishes plus the required back-off interval. The other way round, a device that already transmitted its packet and falls back into low-power listening mode will be waked up by the replies of the remaining devices. Although the transceiver's logic discards those packets very quickly due to a mismatching address, the transceiver still spends much more time in active receive mode.

With an increasing number of concurring replies, the energy consumption of all devices raises to 1030 mJ in average for 38 active devices, which is an increase by factor 18. In addition, the deviation of the energy consumption increases as well. This is caused by the varying time instant, where the device finally transmits its packet. If a device sends its packet very late by repeatedly hitting a large back-off interval during the contention phase, its transceiver continuously stays in an active receive mode to keep track of the ongoing transmissions. If a device transmits its packet very early, it can at least repeatedly fall back into low power listening mode and save more energy than in the first case.

The results of this experiment show, that the standardized ETSI collision avoidance algorithm may conform to the requirements of a logistics scenario in terms of sufficient throughput, but the large impact on the energy consumption of all devices in the network cannot be neglected for this use case. Hence, the distributed channel access brings great potential for optimizations. For example, devices could send their replies on a different channel and use a blind back-off mechanism, which shuts down the receiver during the back-off period. We will elaborate and evaluate those approaches in future works with the help of PhyNetLab.

## X. Conclusion

In this paper we presented PhyNetLab, an IoT testbed which enables a real-life evaluation of future smart logistic approaches. This testbed is built in a materials handling research facility, where all the components for materials handling are already available. This becomes an ideal test field for emulating picking applications and sorting applications that are typical for a materials handling facility. A picking process with smart containers would require random medium access with bi-directional low latency data transmission. In addition, the dynamics of the radio communication differ from a lab environment. Therefore, it is necessary to implement a testbed very close the real world scenario and investigate the dynamics of PhyNodes' communication in such an environment, where the nodes are attached to industrial containers and are used in the logistics facility.

Moreover, the scope of this paper focused on the evaluation of common channel access approaches for radio communications in such an industrial IoT deployment. For this purpose, we implemented a radio-interface driver, which conforms to the ETSI specification for clear channel assessment of Short Range Devices (SRD). The driver is integrated in a novel embedded operating system *Kratos*, which highly customisable and includes efficient mechanisms for energy management and energy accounting of distinct peripheral components. Corresponding energy models are automatically generated once during development and can be included into the deployed system for a live in-system accounting.

On this basis we performed a throughput and energy analysis of a realistic logistics application in the presented testbed, which is polling for a particular product in a warehouse. We showed, that synchronous replies of many PhyNodes to a single poll challenge the channel access algorithm and the energy demand of the entire network. Although the packet loss rate reaches 50 % at high numbers of replying PhyNodes, it is still acceptable for this use case, as long as the number of replies satisfies the demanded amount of goods. But the resulting *storm* of numerous reply packages leads to a significant increase of energy consumption of every single PhyNode in the network. This is caused by frequent wake ups of the radio chip due to radio channel activity and a larger listening period while waiting for a clear channel assessment. Therefore, such deployments as smart warehouses cannot rely on commonly established approaches for radio communication, but rather need specialised solutions for this application.

In future work we will incorporate the PhyNetLab to develop and validate more energy-efficient communication protocols, which satisfy the the demands for scalability, extreme energy-efficiency, and a reasonable latency.

ACKNOWLEDGMENT

Part of the work on this paper has been supported by Deutsche Forschungsgemeinschaft (DFG) within the Collaborative Research Center SFB 876 "Providing Information by Resource-Constrained Analysis", project A4.